# The road to the Loulan Kingdom

## Amelia Carolina Sparavigna
**Department of Applied Science and Technology, Politecnico di Torino**

*Here a discussion on the Loulan Kingdom, an ancient kingdom along the Silk Road, and some observations based on satellite images.*

Since the Antiquity, silk was precious and then travelled westwards from the Far East. The Egyptians and then the Romans began to use it as early as the VIII century BCE. The history of the Middle Ages reveals that monks firstly brought some cocoons to Byzantium from China and that the silk and cocoons trade spread from Byzantium to the Western Europe during the 7th Century. The roads on which this lucrative trade moved got the name of Silk Road.

The Silk Road was an ancient network of trading routes linking China to the West, starting from the Xi'an  The route was composed by caravan routes that played an important part in the exchange of goods but also information between the civilizations of East and West (Waugh, 2009). One most important points of convergence of this network was Kashgar, at the West end of the Taklamakan Desert and the Tarim Basin. The caravans of merchants then followed the road leading to the Caspian Sea, passing through the Afghan valleys, or they climbed the Karakorum Mountains and passed through Iran. Marco Polo took the Silk Road to reach China.

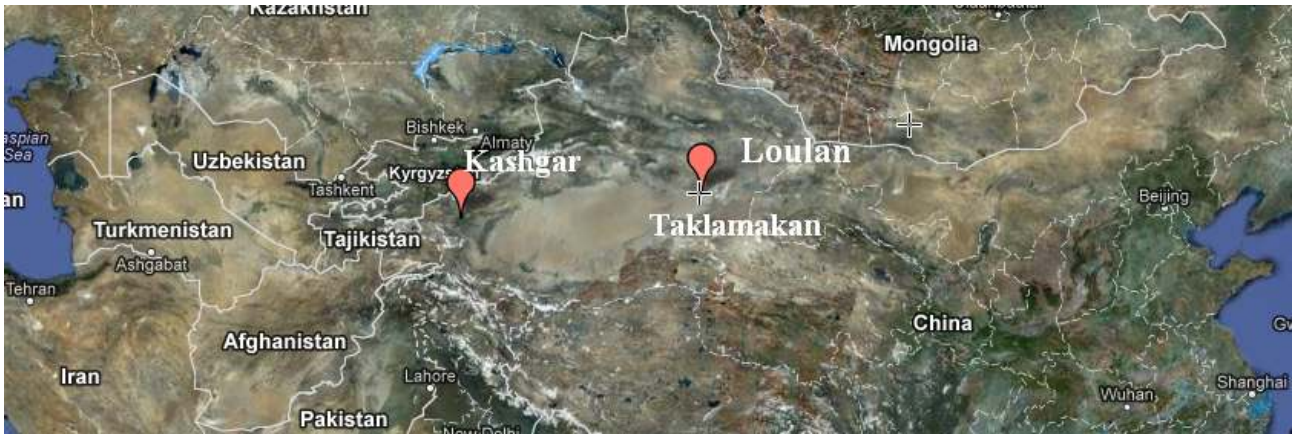

Figure 1 - A huge sea of sand is in the middle of Asia.
This is the Taklamakan desert, as seen in the Google Maps.

We can imagine the huge desert region of the Taklamakan, full of sand dunes, as a sea the caravans had to cross during their long and strenuous journeys. Along the Silk Road, at the East end of the Tarim Basin, there was an ancient kingdom, the kingdom of Loulan, based around an important oasis on the edge of the Lop Desert.

Loulan was already known in the second century BCE. In 108 BCE, the army of the Han Dynasty defeated that of the Loulan kingdom, making it an allied state. In 77 BCE, the Loulan king, Chang Gui, was assassinated and the kingdom came under the control of the Han empire. The ruins of the town of Loulan are on the western banks of Lop Nur basin, now completely desiccated. In fact, Loulan was abandoned after it became completely surrounded by the desert and the region subjected to several attacks by nomadic populations.

Loulan was on the main route from Dunhuang to Korla, where this branch of the Silk Road joined the so-called "northern route", and was also connected by a route due southwest to the town of Wuni in the Charkhlik/Ruoqiang oasis, and, from this place, to Khotan and Yarkand.

The first historical mention of Loulan was in a letter from the Chanyu of the Xiongnu to the Chinese Emperor in 126 BCE in which he claimed that he conquered the Yuezhi, the Wusun, Loulan, and Hujie, "as well as the twenty-six states nearby."

In 126 BCE, the Chinese envoy, Zhang Qian described Loulan as a fortified city near Lop Nur. Because of its strategic position on what became the main route from China to the West for a long period, the control of Loulan was regularly disputed between China and the Xiongnu, which were ancient nomadic-based people that formed a state or confederation north of the empire of the Han territory.

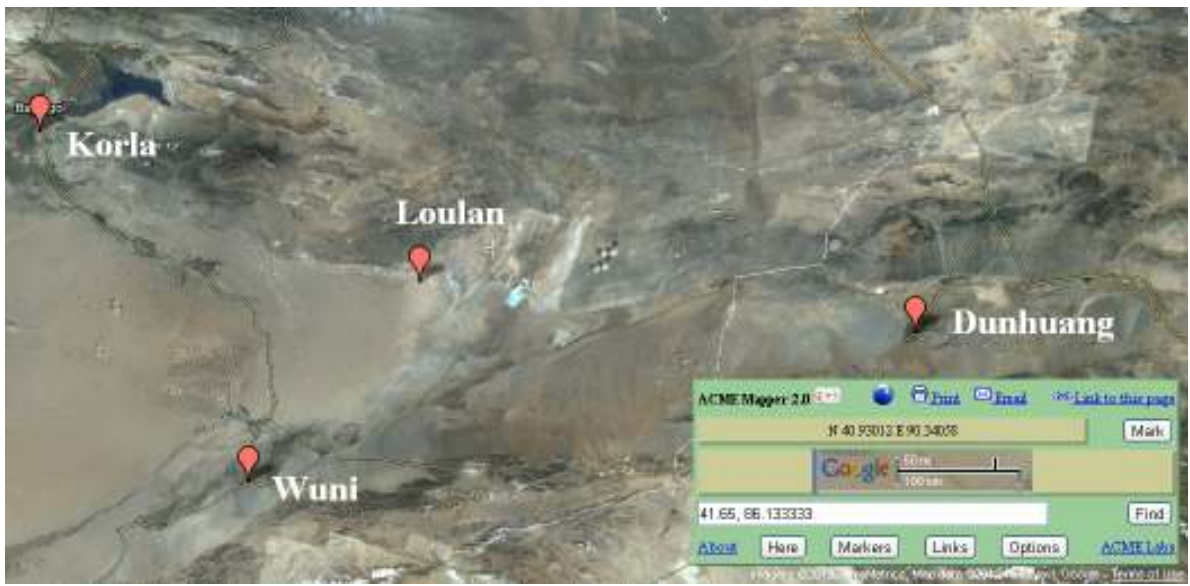

Figure 2 - Here the position of the town of Loulan and of other important towns on the Silk Road.

Here in the following an example of relationships of Loulan with Xiongu and Han Dynasty form the Book of Han, a book reporting the history of China under the Western Han from 206 BC to 25 AD, abstracted from Wikipedia.

In the second century BCE, the Emperor Wu of Han would like extending contact with Fergana. However contacts were made difficult by the Loulan people. Loulan was therefore attacked in 108 and its king captured, after which Loulan agreed to pay a tribute to Han. The Xiongu people, after had become aware of these events, attacked Loulan too. The king of Loulan therefore decided to send one of his sons as a hostage to Xiongnu and another to the Han court. Of course, this behavior upset the Chinese court: the king of Loulan was taken and interrogated, but he answered that Loulan was a small state lying between large states. It seems that he succeed to satisfy the Han emperor concerns because he was released.

The king of Loulan died in 92 BCE, and his countrymen requested the Han court that the king's son be returned. However, the Han court had castrated him and so refused this request, telling that "the Han Emperor had grown too fond of him to let him go". In Loulan, another king was crowned, and again a son was sent as a hostage to the Han court. After the death of this king, the Xiongnu returned the hostage son, named Chang Gui back to Loulan to rule as king. The Han, becoming aware of this fact, demanded that the new king presented himself to the court. Chang Gui refused, considering the fact that the hostages sent to the Han court had been never released. We can imagine that the Loulan king included amongst his concerns the treatment suffered by the first hostage too.

In 77 BCE, a Chinese named Fu Jiezi was sent to kill the Loulan king after several Han envoys were killed or kidnapped. Under the pretext of a present to the king, Fu Jiezi stabbed the Loulan king to death while he was drunk. The king's younger brother Wei-tu-qi was crowned king of Loulan by the Han court, and the kingdom was renamed Shanshan under the Han control. The new king asked for some Han forces be established nearly, due to his fear of revenge from the sons of the assassinated king. Chinese soldiers were therefore sent to occupy this area.

It seems that about a thousand of Chinese soldiers were established at Loulan in 260 CE. The site was abandoned in 330 CE due to lack of water when the Tarim River, which supported the settlement, changed its river bed (Zhibao et al.,2012; Zhang et al., 2003, Yuan et al., 1999). From the 5th century, the land was frequently invaded by nomads and the area became gradually abandoned: at the beginning of the Tang Dynasty period the scattered Shanshan people migrated to a Northern area and the Loulan region was completely depopulated (Yuang et al., 1999). A Buddhist pilgrim, named Xuanzang, who passed through this region in 644 on his return from India to China, wrote, "A fortress exists, but not a trace of man". Probably, this is the fort of Yingpan to the northwest of Loulan that remained under Chinese control until the Tang Dynasty. And the Buddhist pilgrim Faxian who passed in Shanshan in 399 on his way to India, described "a country rugged and hilly, with a thin and barren soil. The clothes of the common people are coarse, and like those worn in our land of Han".

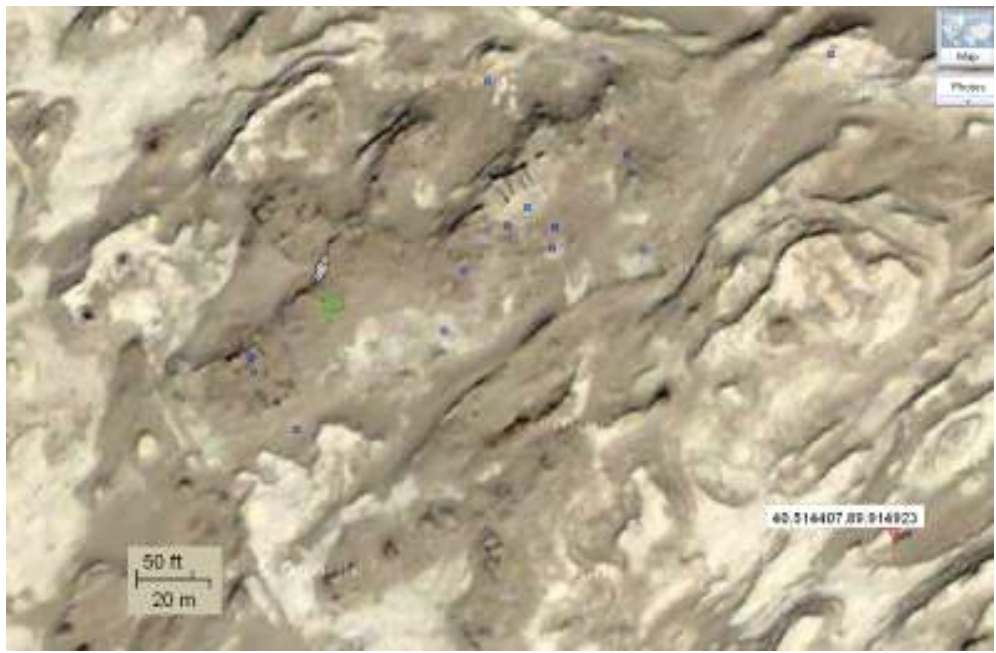

Figure 3 - The archaeological site of the Loulan town as seen in the satellite maps (Google Maps). The ruins are hardly detectable on this arid land. Selecting one of the blue dots it is possible to see pictures uploaded by people who visited the place.

Archaeology tells that the kingdom of Loulan/Shanshan had a walled town near the northwest corner of Lop Nur, near Tarim River, before the river changed its bed. The ruin city of Loulan was discovered by Sven Hedin in 1900, who found also many Chinese manuscripts from the Western Jin Dynasty (265–420). Aurel Stein made further excavations in 1906 and 1914. Stein found a wool-pile carpet fragment, silk, and Gandharan architectural wood-carvings. In 1979 and 1980, three archaeological expeditions sponsored by the Chinese Academy of Social Sciences performed excavations in Loulan. These expeditions discovered a manmade canal, running through Loulan from northwest to southeast; a Buddhist stupa and a large home apparently for a Chinese official.

They also collected several vessels of wood, bronze objects, jewelry and coins, and Mesolithic stone tools. More recently, a Chinese expedition has discovered several other archaeological places in this area (Lü et al., 2010).

Probably, the early settlers in this area were Tocharians, some Indo-European people. In fact, the Tarim basin is also known for the Tarim mummies, a series of mummies, which date from 1900 BCE to 200 CE. Some of these mummies are frequently associated with the presence of the Tocharian people (Mair, 2010; Coppens, 2009; Pringle ,2010).

As we can see from the satellite maps, such as the Google Maps, the town of Loulan is hardly detectable: we can see a few structures arising in the desert. Using a larger scale, we see an aeolian sandy soil (Wu et al.,2009) striated in a SW-NE direction, by the blowing of dominant winds. About 20 km far from the site of the Loulan town, satellite maps reveal a huge structure, having walls 100 meters long (see Fig.4). It seems the remain of a caravanserai, a roadside inn where travelers could rest and recover from the day's journey. Caravanserais supported the flow of commerce in particular along the Silk Road. From the image, it is clear that this structure was oriented in such a manner the walls shelter the site from the prevailing winds. Some archaeologists defined it as an ancient city, naming it the LE city (Lü et al., 2010).

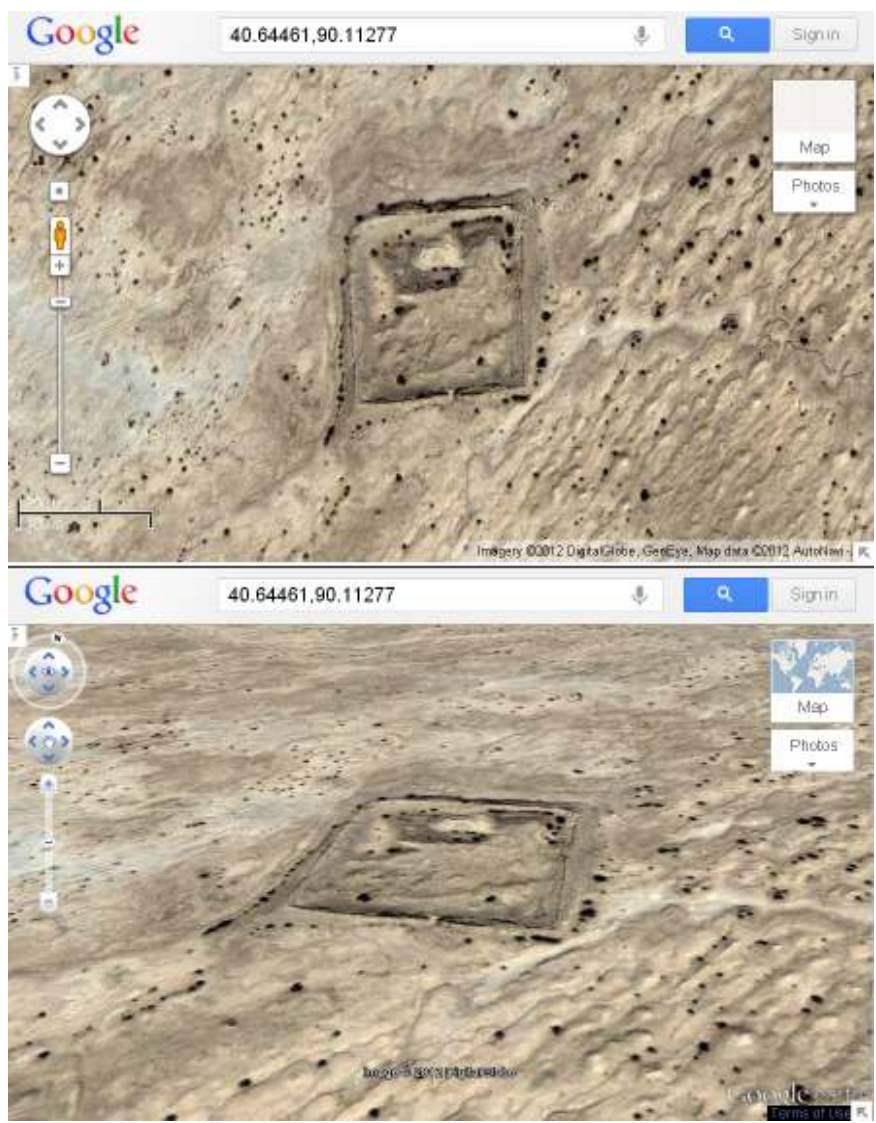

Figure 4 - 20 km from Loulan, a huge structure with walls 100 meters long can be seen in the desert. Some archaeologists consider it as a city, calling it the LE city (Lü et al., 2010). It looks like a caravanserai, a roadside inn where travelers could along the Silk Road.

The caravanserai in the Figure 4 is probably a reference point for the modern travelers, which want to arrive in Loulan; in the Acme Maps we can see an almost straight track starting from it (origin O in the Figure 5). Some markers can be useful when following this track. Some faint tracks, perpendicular to this one can be detected in the satellite maps. One of them is passing near Loulan.

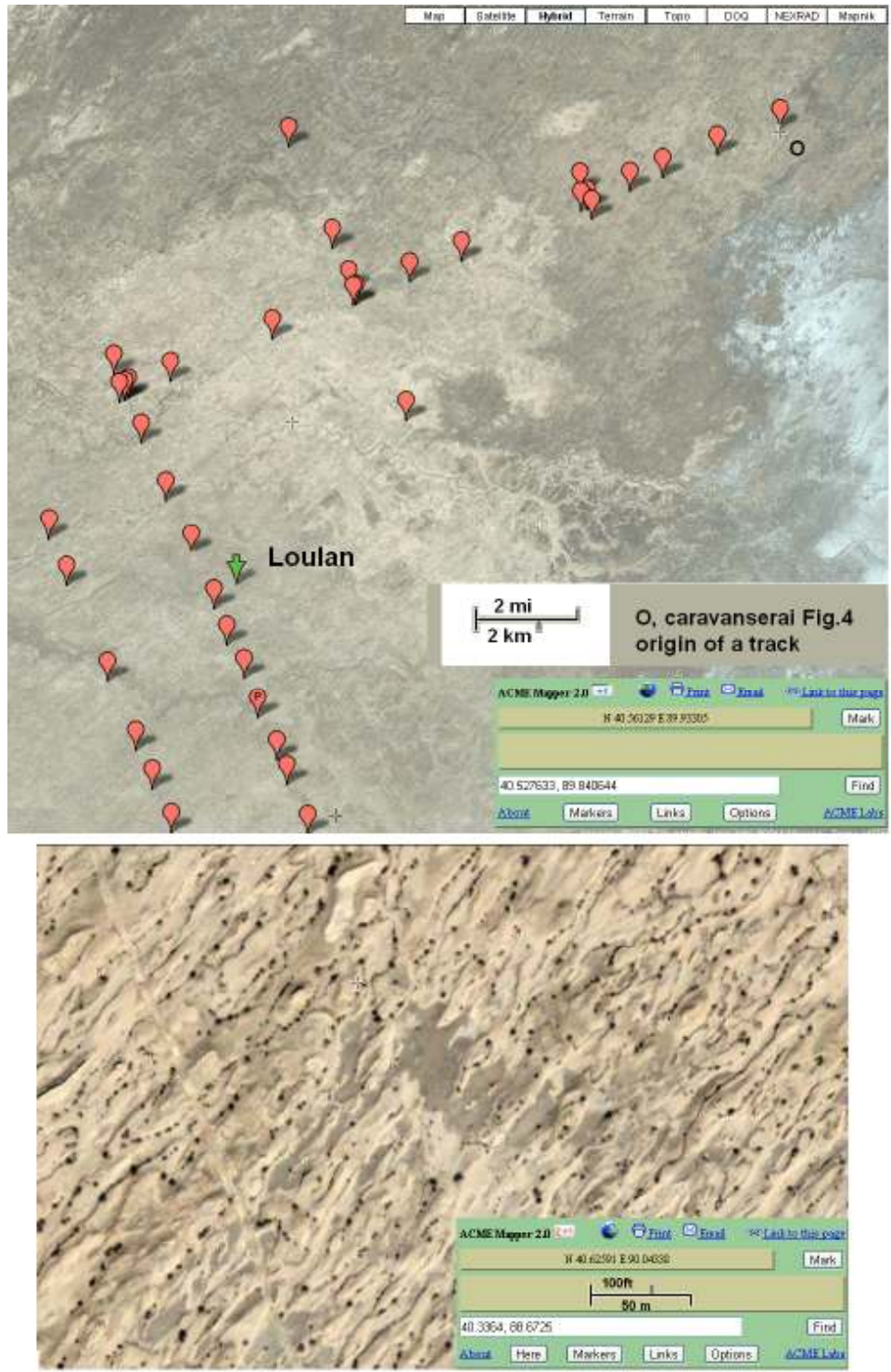

Figure 5 - Tracks in the desert near Loulan. One is originating from the structure shown in the Figure 4. Others are perpendicular to this track. In the lower part of the figure, a detail of one of the tracks. It is visible on the left of the image crossing the striated texture of the land. The track is not North-South oriented, but inclined of about 20 degrees. Probably, this is a more convenient manner to cross such an aeolian sandy soil, instead of moving on the precise North-South direction.

As previously told, archaeological expeditions has discovered an artificial canal, running through Loulan from northwest to southeast. May be, the tracks that we see in the satellite images are partially following some ancient canals too. After the Tarim River diverted its flow, these tracks could have be used to move in the desert. It is quite probable that they are recent tracks to go across this desert, used by the archeologists to survey it. These tracks are not North-South oriented, but are inclined of about 20 degrees. Using a freely available planetarium for PC, the Stellarium, we can see that two thousands of years ago, if we were looking in this precise NNW direction, we could have seen the setting of Cassiopeia, and in particular of the five stars forming a 'W' shape, which is the distinctive sign of this constellation. Thinking of similar tracks in ancient times, we could imagine some travelers of the Silk Road using them to move their caravans, aiming the setting of some stars.

In the Google Maps, it is possible to observe some textures (Fig.6), different from that of the surrounding soil. These could be some ancient manmade structure, as that shown in Fig.4.

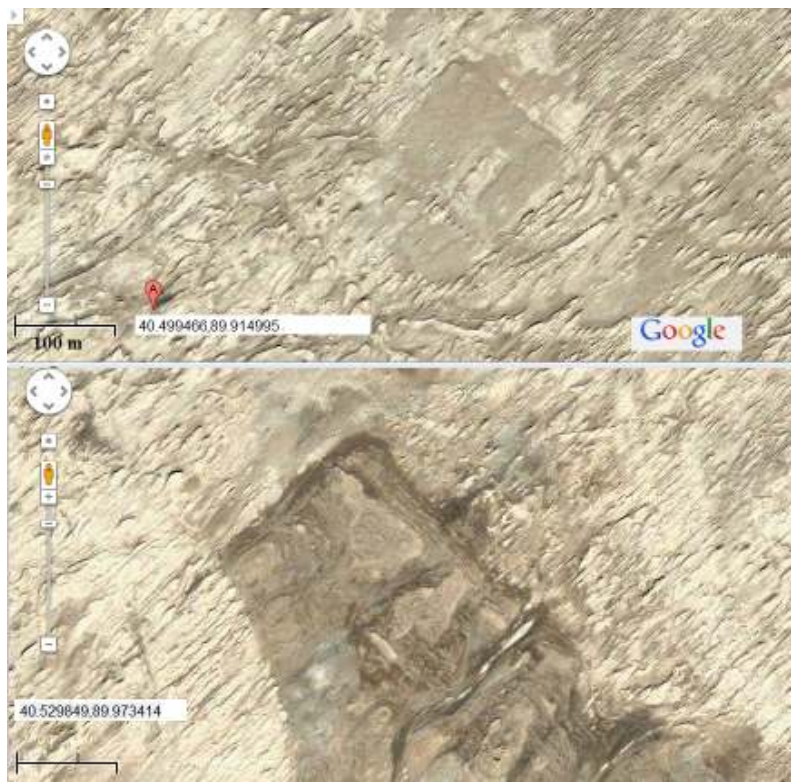

Fig.6  Probably manmade structures near Loulan.

The history of Loulan was strongly conditioned by the environmental changes. In particular the Tarim river, changing its bed, eventually caused the inhabitants to abandon the region. Recently, many geophysical researches (Zhibao et al.,2012; Hongjuan et al. 2012) had been devoted to study this area. Zhang et al. (2003) deeply discussed the evolution of the Tarim Basin, in particular of the oases. The researchers considered the climate change and the role of human activities too. They are remarking that the population growth in this region is leading to a rapid expansion of cultivated land, which needs water for irrigation. As a result, the utilization of the natural resources in these oases produces increasingly destructive effects on their natural environments.

In many of the references on the Loulan environment, the satellite images reveal their usefulness in characterizing the different textures of the soil. Besides the geophysics researches, they help finding some archaeological remains. If the satellite maps had enough resolution, we could follow and investigate the network of the Silk Road and find some ancient sites along it, lost in the desert sand.